\documentclass[prb, aps, superscriptaddress, amsmath, amssymb, preprint, 12pt]{revtex4-2} 

\usepackage[english]{babel}

\usepackage{graphicx}
\usepackage{siunitx}
\usepackage{xcolor,xspace,colortbl,rotating}
\usepackage{nicefrac}
\usepackage{academicons}

\usepackage{setspace}
\usepackage[a4paper,left=2.4cm,right=2.4cm,top=2.4cm,bottom=2.4cm]{geometry}

\newcommand{\dd}{\mathrm{d}}

\renewcommand{\vec}[1]{\bm{#1}}        

\newcommand{\muS}{\mu_S}

\newcommand{\kBT}{k_\mathrm{B}T}
\newcommand{\Neel}{N\'{e}el}

\begin{document}
	
	\pagestyle{empty}
	
	\title{{Exceptional sign changes of the non-local spin Seebeck effect in antiferromagnetic hematite}}
	
	\author{Andrew \surname{Ross}}
	\altaffiliation[Current Affiliation: ]{Unit\'e Mixte de Physique CNRS, Thales, Universit\'e Paris-Saclay, Palaiseau 91767, France}
	\affiliation{Institute of Physics, Johannes Gutenberg-University Mainz, 55128 Mainz, Germany}
	\affiliation{Graduate School of Excellence Materials Science in Mainz, 55128, Mainz, Germany}

	\author{Romain \surname{Lebrun}}
	\affiliation{Unit\'e Mixte de Physique CNRS, Thales, Universit\'e Paris-Saclay, Palaiseau 91767, France}
	
	\author{Martin \surname{Evers}}
	\affiliation{Fachbereich Physik, Universität Konstanz, 78457 Konstanz, Germany}

	\author{Andr\'as \surname{De\'ak}}
	\affiliation{Department of Theoretical Physics, Budapest University of Technology and Economics, H-1111 Budapest, Hungary}

	\author{L\'aszl\'o \surname{Szunyogh}}
	\affiliation{Department of Theoretical Physics, Budapest University of Technology and Economics, H-1111 Budapest, Hungary}
	\affiliation{MTA-BME Condensed Matter Research Group, Budapest University of Technology and Economics, Budafoki út 8, H-1111 Budapest, Hungary}

	\author{Ulrich \surname{Nowak}}
	\email{ulrich.nowak@uni-konstanz.de}
	\affiliation{Fachbereich Physik, Universität Konstanz, 78457 Konstanz, Germany}

	\author{Mathias \surname{Kl\"aui}}
	\email{klaeui@uni-mainz.de}
	\affiliation{Institute of Physics, Johannes Gutenberg-University Mainz, 55128 Mainz, Germany}
	\affiliation{Graduate School of Excellence Materials Science in Mainz, 55128, Mainz, Germany}
	\affiliation{Center for Quantum Spintronics, Department of Physics, Norwegian University of Science and Technology, NO-7491 Trondheim, Norway}
	
	\date{\today}
	
%
%
	\begin{abstract}
		\textbf{Low power spintronic devices based on the propagation of pure magnonic spin currents in antiferromagnetic insulator materials offer several distinct advantages over ferromagnetic components including higher frequency magnons and a stability against disturbing external magnetic fields. In this work, we make use of the insulating antiferromagnetic phase of iron oxide, the mineral hematite $\alpha$-Fe$_2$O$_3$ to investigate the long distance transport of thermally generated magnonic spin currents. We report on the excitation of magnons generated by the spin Seebeck effect, transported both parallel and perpendicular to the antiferromagnetic easy-axis under an applied magnetic field. Making use of an atomistic hematite toy model, we calculate the transport characteristics from the deviation of the antiferromagnetic ordering from equilibrium under an applied field. We resolve the role of the magnetic order parameters in the transport, and experimentally we find significant thermal spin transport without the need for a net magnetization.}
	\end{abstract}
%
%
	
	\maketitle

%
%
%
	\section{Introduction} \label{sec:Intro}
	The quanta of magnetic excitations, known as magnons, efficiently transport angular momentum in magnetic materials including ferromagnetic \cite{Cornelissen2015, Goennenwein2015} and antiferromagnetic insulators (AFMI) \cite{Lebrun2018, Ross2020, Han2020}.
	Such magnon spin currents can be excited in AFMIs by several external stimuli including an interfacial spin-bias \cite{Lebrun2018, Ross2020, Han2020, Lebrun2020}, magnetic resonance \cite{Vaidya2020, Li2020, Boventer2021room} or a temperature gradient $\nabla T$ \cite{Li2020, Wu2016, Reitz2020, Rezende2018}. 
	This third effect is known as the spin-Seebeck effect (SSE) where the combination of a magnetic field $\vec{H}$ and thermal gradient $\nabla T$ induces a magnonic spin current \cite{Li2020, Wu2016, Reitz2020, Rezende2018, Seki2015, Li2019}. 
	If a heavy metal (HM) with a large spin orbit coupling is placed in contact with the AFMI, this magnon current can be converted to a charge current and thus detected due to the inverse spin Hall effect (ISHE) \cite{Li2020, Wu2016, Lebrun2018, Sinova2015}. 
	However, whether the spin current can be detected has been shown to depend on the symmetry of the antiferromagnetic ordering with respect to both $\nabla T$ and $\vec{H}$ \cite{Rezende2018, Wu2016, Seki2015}. \\
	
\begin{figure}[ht!]
		\includegraphics[width = 8cm]{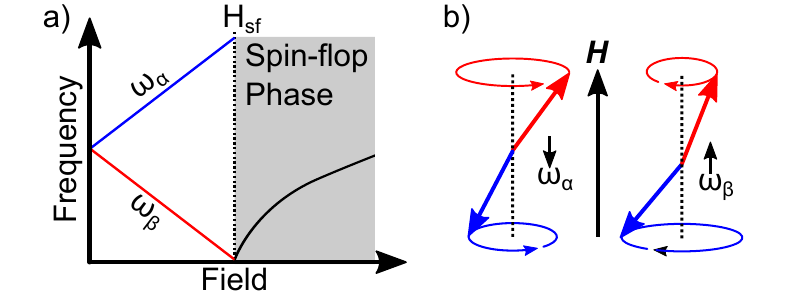}
		\caption{\label{fig:magnon modes}a) Schematic magnon dispersion as a function of an applied field for a uniaxial antiferromagnet. The application of a magnetic field splits the degeneracy of the two lowest magnon modes. At the spin-flop field $H_{sf}$, the frequency of one mode reaches zero and the two circular modes change character. b) Schematic of the two easy-axis modes. The right circularly polarized mode $\omega_\alpha$ and the left circularly polarized mode $\omega_\beta$.}
	\end{figure}

	The generation of a magnonic spin current by a thermal gradient offers an attractive alternative for magnonic applications and even the local, active manipulation of magnon currents excited by other mechanisms \cite{Cramer2019}. 
	Furthermore, using antiferromagnetic materials as magnonic spin current conduits offer several benefits over ferromagnetic materials including a resilience to external disturbing magnetic fields and higher frequency magnon excitations \cite{Baltz2018}.
	Bilayers of AFMI/Pt have previously been employed to investigate the locally generated voltage in response to an interfacial temperature gradient applied out of the sample plane, perpendicular to the antiferromagnetic anisotropy direction and the detected transversal voltage \cite{Seki2015, Wu2016, Shiomi2017, Li2019, Li2020}. 
	The used AFMIs have a uniaxial anisotropy, such that a spin-flop, where the antiferromagnetic \Neel{} vector $\vec{n}$ spontaneously rotates \SI{90}{\degree}, is induced for a magnetic field applied parallel to the anisotropy easy-axis. 
	While in some systems, a signal below the spin-flop field was observed and in others not, the consensus is that the spin-flop has a strong impact on the local SSE signal. 
	These measurements are explained by making use of a magnon spin diffusion model based on degenerate magnon modes. 
	This degeneracy is broken in the presence of a magnetic field, leading to a net imbalance between the two lowest, $k=0$ modes and thus a detectable magnon current \cite{Rezende2018}. 
	So far for the thermally driven spin current, no high field suppression associated to magnon suppression (as seen in the ferrimagnetic insulator Y$_3$Fe$_5$O$_{12}$(YIG) \cite{Cornelissen2016}) has been observed even above the spin-flop field where an increasing field serves to increase the magnon gap. 
	It has been suggested that the thermally induced voltage is directly related to the net magnetization \cite{Seki2015, Li2019} but this is disputed for instance for the AFMI $\alpha$-Cu$_2$V$_2$O$_7$ \cite{Shiomi2017}, suggesting that the AFM-SSE depends not only on the magnetic properties but also the transport properties of the investigated AFM, where the magnon relaxation time dominates the transport, particularly at low temperatures. 
	Crucially, these previous studies have all found a strong suppression of the local SSE with increasing temperature.
	At low temperatures, a peak in the local SSE voltage has been reported, explained via the competing contributions of the two lowest $k=0$ magnon modes \cite{Rezende2018}.
	The increasing temperature leads to an increase in the population of the lowest, $k=0$ magnon mode [see Fig. \ \ref{fig:magnon modes}], until the thermal energy overcomes the magnon gap of the higher energy mode. 
	These two magnon modes contribute to the net observed voltage with opposing signs, and thus a maximum occurs just below the temperature required to overcome the magnon gap of the higher energy $k=0$ mode \cite{Rezende2018}.
	
%
%
%
%
	
	The application of the magnetic field perpendicular to $\vec{n}$, such that the magnetic sublattices cant in the direction $\vec{H}$, leads to a net magnetization.
	This configuration has also been shown to elicit a response \cite{Seki2015, Li2019} but the importance of the net magnetization for the transport is disputed \cite{Wu2016, Shiomi2017}. 
	Moving away from local thermal gradients across the interface, the non-local SSE (NL-SSE) has also very recently been investigated in AFMIs where $\nabla T$ now exists between two non-magnetic metal electrodes \cite{Lebrun2018, Yuan2018, Xing2019, Hoogeboom2020, Rezende2018, Muduli2021, Bender2017, Hoogeboom2021}. 
	It was found that a large field induced magnetization parallel to the transport direction was required to enable significant spin transport of thermally excited magnons \cite{Lebrun2018, Yuan2018, Xing2019, Hoogeboom2020}.
	An increase in temperature led to a sharp suppression of the NL-SSE, just as for the local signal \cite{Hoogeboom2020, Yuan2018, Xing2019, Muduli2021}, however, a significant signal was still reported at \SI{200}{\K} for $\alpha$-Fe$_2$O$_3$ \cite{Lebrun2018}.
	 {Very recently, it was also suggested that the NL-SSE in antiferromagnets may be super-imposed by the transverse spin Nernst effect \cite{Wimmer2021}, however for the work performed here, the transverse spin Nernst effect can be excluded as a major effect due to geometrical considerations. } \\
	
	The transported spin current due to a thermal gradient is given by $J_S= -(\sigma_m \nabla \mu_m + S \nabla T)$, where $\sigma_m$ is the magnon conductance, $\nabla \mu_m$ is the net magnon population between the different contributing modes and $S$ is the spin Seebeck conductance. 
	Antiferromagnets exhibit two magnetic order parameters, the staggered \Neel{} vector $\vec{n}$ and a net magnetization $\vec{m}$ that can appear under an applied field.
	This then means that $S$ has contributions related to each order parameter; the \Neel{} order spin Seebeck conductance $S_{\vec{n}}$ and the magnetic moment spin Seebeck conductance $S_{\vec{m}}$ \cite{Rezende2018, Reitz2020}. 
	Each of these comes with additional transport and dissipation paths for thermally excited magnons and has a distinct field and temperature dependence \cite{Rezende2018, Reitz2020}.
	Generally, based on the observations in both local and non-local spin Seebeck measurements, $S_{\vec{m}} \gg S_{\vec{n}}$ so the key the question is of whether $S_{\vec{n}}$ can enable significant spin transport over long-distances at elevated temperatures. \\
	
	In this work we demonstrate that the application of a magnetic field parallel to the easy-axis of  {a bulk crystal of} the insulating antiferromagnetic iron oxide hematite ($\alpha$-Fe$_2$O$_3$) elicits a non-local spin Seebeck effect  {mediated by the N\'eel vector}. 
	Magnons are efficiently transported both parallel and perpendicular to the applied field with distinct dependencies and we show that the applied field breaks the degeneracy of the available magnon modes facilitating a net transport along the thermal gradient direction. 
	 {Finally, we develop a toy model of thermal transport in hematite, where we are able to resolve some of the key experimental observations.}
	We utilize $\alpha$-Fe$_2$O$_3$ because it has been shown to facilitate long-distance magnon transport excited by in interfacial spin-bias \cite{Lebrun2018, Lebrun2020, Ross2020, Han2020}, a low magnetic damping \cite{Lebrun2020, Fink1964} and a controllable magnetic order in bulk crystals and thin films \cite{Lebrun2019, Morrison1973, Ross2020b}. 
	Below the Morin temperature \cite{Morin1950}, this material adopts an easy-axis (EA) anisotropy where the \Neel{} vector $\vec{n}$ aligns parallel to the crystallographic c-axis \cite{Lebrun2019,Morrison1973}.
	An antisymmetric exchange interaction, the Dzyaloshinskii-Moriya interaction (DMI), parallel to the c-axis is present in this material and leads to a reorientation of $\vec{n}$ under an applied field if a finite angle between $\vec{H}$ and the EA exists \cite{Lebrun2019, Morrison1973}. 
	
%
%
	\section{Thermal Magnon Transport Along the Easy Axis}\label{sec:transport para}
\begin{figure*}[ht!]
		\includegraphics[width=12.9cm]{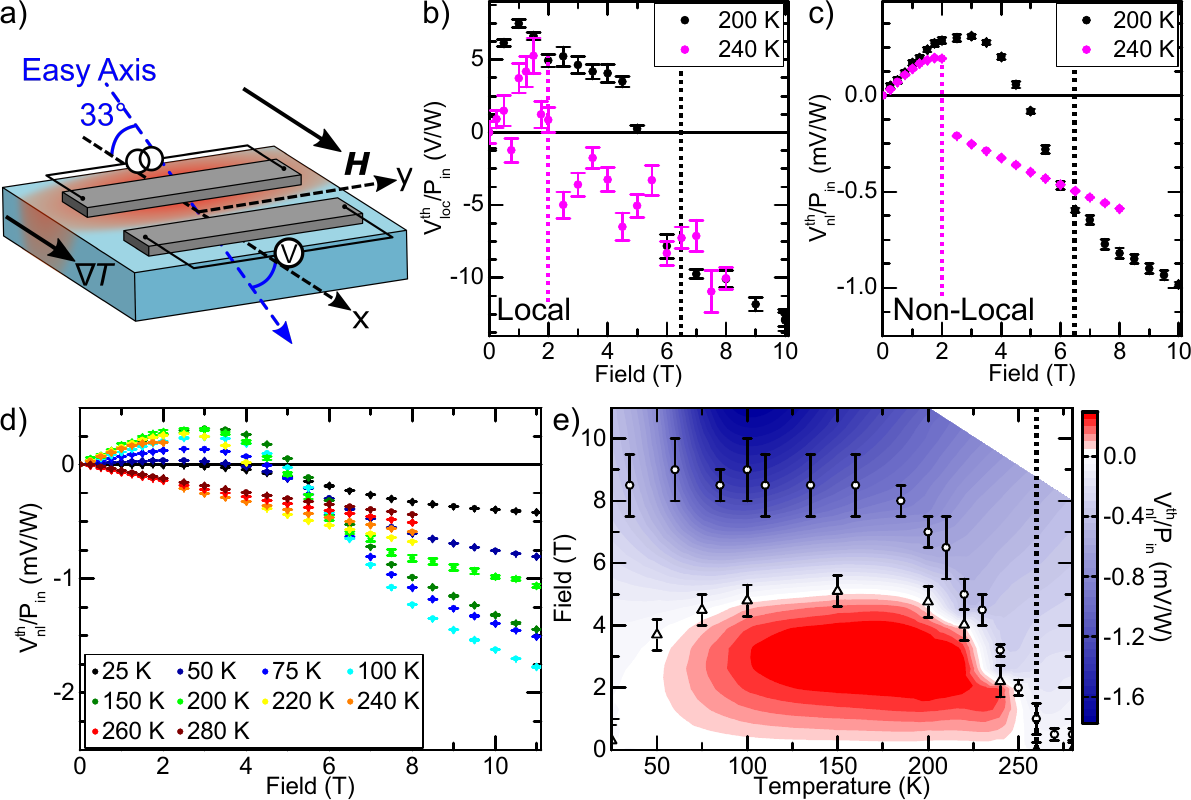}
		\caption{\label{fig:Wires Perp}a) Geometry used to investigate the non-local spin Seebeck effect (NL-SSE) in hematite (blue). A passing charge current heats the hematite below one Pt wire, generating a thermal gradient $\nabla T$. The easy-axis is inclined at \SI{33}{\degree} to the surface plane and a magnetic field $\vec{H}$ is applied parallel in the in-plane component of the \Neel{} vector. b) Local SSE as a function of magnetic field at $T=\SI{240}{\K}$ (magenta) and $T=\SI{200}{\K}$ (black). The dashed line indicates the spin-flop field at the respective temperature. c) NL-SSE at $T = \SI{240}{\K}$ (magenta) and $T = \SI{200}{\K}$ (black) as a function of the applied magnetic field. The dashed lines indicate the spin-flop field at the respective temperature. d) NL-SSE as a function of the applied magnetic field parallel to the in-plane component of the easy-axis for temperatures above and below the Morin transition temperature ($T_M = \SI{257}{\K}$). e) NL-SSE as a function of field and temperature. Open circles indicate the spin-flop field from Ref.\ \protect\cite{Lebrun2019} and triangles indicate the inversion field from d). The dashed line indicates $T_M$.  {The magnitude of $\vec{H}$ stated is that along $\vec{x}$.}}
	\end{figure*}

	 {We start by investigating the efficiency of magnon transport in $\alpha$-Fe$_2$O$_3$ parallel to the easy-axis. Electrically excited magnons have been shown to be efficiently transported by the N\'eel vector \cite{Lebrun2018, Ross2020}, but so far, the N\'eel spin Seebeck coefficient has been shown to be too small to elicit significant signals \cite{Hoogeboom2020, Lebrun2018, Xing2019, Yuan2018}.}
	Using lithographic methods \cite{Lebrun2018, Ross2020}, we define a non-local geometry of \SI{7}{\nm} Pt wires, \SI{250}{\nm} wide and \SI{80}{\micro\m} long, separated by \SI{500}{\nm} atop a commercially obtained $\alpha$-Fe$_2$O$_3$  {bulk} crystal, orientated as (1$\bar{1}$02) such that the easy-axis is inclined \SI{33}{\degree} to the surface plane [see Fig. \ \ref{fig:Wires Perp}a)]. 
	A charge current is applied to one wire, which leads to Joule heating of the underlying magnetic material, generating a thermal gradient between the two wires, and the excitation and flow of thermally excited magnons due to the SSE \cite{Cornelissen2015, Flebus2019, Rezende2018, Bender2017}. 
	The second Pt wire acts as a detector, absorbing the flowing spin current underneath, converted to a measurable voltage by the ISHE. 
	The non-local SSE (NL-SSE) voltage is defined as $V^{th}_{nl} = \left(V_{nl}^+ + V_{nl}^- \right)$ where $V_{nl}^+$ ($V_{nl}^-$) is the non-local voltage detected for a positive (negative) current $I$, where we invert the current to remove any offsets and isolate the thermal signal. 
	$V^{th}_{nl}$ is then normalized to the input power $P_{in} = R \cdot I^2$ where the resistance of the injector $R$ takes into account any change in the resistance due to the spin Hall magnetoresistance (SMR) \cite{Lebrun2019, Manchon2017, Baldrati2018, Hoogeboom2017, Fischer2018, Geprags2020}.
	 {During the fabrication process, the Pt/$\alpha$-Fe$_2$O$_3$ interface is subject to a baking step exceeding \SI{400}{\kelvin}, higher than any Joule heating we may reach during our experiments, allowing us to exclude any impact on Joule-heating-driven annealing of the interface \cite{Shan2016, Kohno2021}.}
	We fix the orientation of an external field $\vec{H}$ parallel to the in-plane projection of the easy-axis,  {such that $\vec{H}$ is at an angle of \SI{33}{\degree} to the easy-axis} [see Fig. \ \ref{fig:Wires Perp}a)].
	Below $T_M$ in the easy-axis phase, this will induce a spin-reorientation at some critical magnetic field $H_{cr}$ where $\vec{n}$ will reorientate perpendicular to $\vec{H}$ \cite{Morrison1973, Lebrun2019}. 
	Above $H_{cr}$, a field induced magnetization will appear directed along $\vec{H}$  {and a net moment will appear out of the sample plane} [see also Appendix \ \ref{appendix:Model Details}].\\
%
%
%
%
	
	We first adopt the geometry shown in Fig. \ \ref{fig:Wires Perp}a) where the thermal gradient $\nabla T$, transport direction and magnetic field $\vec{H}$ are coincident. 
	As well as the thermal gradient between the wires, there will also be a secondary thermal gradient across the Pt/$\alpha$-Fe$_2$O$_3$ interface that can lead to a local SSE \cite{Wu2016, Seki2015, Shiomi2017}.
	The thermally excited voltage due to the local SSE is shown in Fig. \ \ref{fig:Wires Perp}b) as a function of the applied magnetic field for several temperatures below $T_M$. 
	Given the nature of the excitation, we cannot ascertain the precise thermal gradient across the interface, but we can resolve the measured voltage as a function of the input power. 
	If the deposited power is increased, the local signal remains the same (see Appendix \ref{appendix:Current Dependence}), although it can change the nature of the non-local signal discussed later.  
	Close to $T_M$, the signal is noisy, likely due to the thermal fluctuations that accompany the transition around this temperature but the net trend is a small signal below $H_{cr}$ that changes sign and increases in amplitude linearly above $H_{cr}$. 
	Despite the thermal fluctuations, this observation of a finite local SSE also occurs at the highest temperature so far reported for an antiferromagnet \cite{Wu2016, Seki2015, Shiomi2017, Li2019, Li2020, Hoogeboom2020}.
	Further below $T_M$, the signal below $H_{cr}$ more defined due to the reduced thermal fluctuations and also changes sign in the vicinity of $H_{cr}$ and the amplitude is linear by increasing with increasing magnetization.
	 {If we consider the individual components of $\vec{n}$ and $\vec{m}$ for these local measurements [see Appendix Sec. \ref{appendix:Model Details}, Fig. \ref{fig:equilibriumPosition_n and m}], the signal above $H_{cr}$ is dominated by the emerging magnetization as previously reported. Below $H_{cr}$, the field-evolution of $\vec{m}$ does not follow that of our experimental results indicating that the orientation of $\vec{m}$ is not sufficient to drive the signal we observe and instead represents a combination of a local SSE mediated by both the N\'eel vector and the magnetization. } 
	The presence of a finite, local signal below $H_{cr}$ has been reported \cite{Wu2016, Shiomi2017, Li2019, Reitz2020} and explained by differences in population of the contributing magnon modes under an applied field, but only at low temperatures below the thermal energy of the higher energy $k=0$ mode [Fig.\ \ref{fig:magnon modes}a)]. 
	In this work however, the temperature of the surrounding environment will lead to thermal population of many magnon modes, not just the lowest $k=0$ mode. \\
	
	Moving to the non-local transport of thermally excited magnons, the lateral thermal gradient $\nabla T$ and $\vec{H}$ are coincident and we would naively expect little, to no transport in the absence of a significant net magnetization below $H_{cr}$ [see Appendix \ref{appendix:Model Details}]. 
	However, surprisingly, we find an observable and increasing non-local signal below the spin-reorientation field $H_{cr}$ in the easy-axis phase just below $T_M$ at \SI{240}{\K} [Fig. \ \ref{fig:Wires Perp}c)].
	The signal reaches a maximum at $\mu_0 \vec{H}$ = \SI{2}{\tesla}. 
	Just above this field, there is an abrupt inversion of the signal coincident with the spin-flop field at this temperature \cite{Lebrun2019, Morrison1973}. 
	Above $H_{cr}$, the NL-SSE is linear with increasing field, consistent with an emerging field induced magnetization parallel to the transport direction \cite{Lebrun2018, Wu2016, Rezende2018}. 
	 {We note that the finite angle between the magnetic field and the easy-axis leads to a finite magnetic moment along the transport direction $\vec{x}$ [see Fig. \ \ref{fig:equilibriumPosition_n and m}], even below the critical field. One could then attribute the non-local signal observed entirely to this net moment. However, the moment along $\vec{x}$ increases almost linearly with the applied field, showing little change at $H_{cr}$, at odds with our experimental observations. Comparing this to the magnitude of the moment along $\vec{z}$ or the orientation of the N\'eel vector, where significant changes happen, the net trend of the field-induced magnetization is insufficient to explain our experimental results.  }
	The presence of a significant signal below the spin-reorientation indicates that the \Neel{} order spin Seebeck conductance $S_{\vec{n}}$ contributes significantly to the signal \cite{Rezende2018, Li2020, Reitz2020}.
	In the absence of a magnetic field, the two lowest magnon modes contributing to the net transport are degenerate and no net transport is observed.
	The application of a magnetic field breaks this degeneracy (as explained schematically in Fig. \ \ref{fig:magnon modes}a)), and net transport is conceivable even without a net magnetic moment providing that $S_{\vec{n}}$ is reasonably large \cite{Rezende2018}. \\
	
	By cooling further below $T_M$, the magnetic field required to induce a spin-reorientation moves to higher magnetic fields, becoming constant at low temperatures \cite{Lebrun2019, Morrison1973}. 
	The NL-SSE at $T = \SI{200 }{\K}$ is shown in black in Fig. \ \ref{fig:Wires Perp}c). 
	Similar to at  \SI{240}{\K}, a significant, increasing signal appears with the application of a magnetic field due to the broken degeneracy of the magnon modes.
	However, contrary to the case of  \SI{240}{\K}, rather than increasing up to the spin-flop field, a turning point appears at $\mu_0$$\vec{H} = \SI{3}{\tesla}$, above which the signal turns negative at $\mu_0$$\vec{H} = \SI{4.7}{\tesla}$ and experiences a change in gradient at the spin-flop field $\mu_0$$\vec{H} = \SI{6.5}{\tesla}$.
	Whilst closer to $T_M$, the behavior was consistent with a more localized SSE as presented by Ref.\ \cite{Reitz2020}, the response at this lower temperature cannot be so easily interpreted.
	 {It was recently discussed by Wimmer \textit{et al.} \cite{Wimmer2020} that the long distance transport of magnons can be achieved through a superposition of arbitrarily polarized magnon modes in an effect akin to a magnon Hanle effect \cite{Kamra2020}. The application of a magnetic field leads to modulation of the magnon frequencies, even leading to a ``beating'' of the magnon transport \cite{Wimmer2020, Ross2020c,Kamra2020}. While it is possible that such an effect could be employed to understand the sign change of the NL-SSE, these previous works have focused on antiferromagnets with a hard-axis anisotropy, where the magnon modes are linearly polarized and excited by a spin-polarized interfacial accumulation. In the case of thermally excited magnons, where there is no selective excitation of the magnon modes, there is no expected modulation of the net response with applied field for easy-axis or hard-axis antiferromagnets \cite{Rezende2018}. Finally, a beating of the magnon transport should also be visible when the distance between the two Pt wires is varied. As we show in the Appendix\ \ref{appendix:Distance Dependence}, we observe the same behavior of the magnon transport, which would only occur if the distances were exact multiples of the beating wavelength, which is unlikely.}
	 {We observe that we can resolve a sharp transition at the spin-flop, when the angle between field and easy-axis is reduced further (see Appendix\ \ref{appendix:Beta Plane}). This demonstrates that, with the complete absence of a net magnetization along the transport direction, that the N\'eel spin Seebeck conductance is capable of transporting a magnon current. Once a finite angle is introduced between the field and easy-axis, this transition softens, and we resolve a turning point indicating that the magnetic and N\'eel spin Seebeck conductances are competing with similar magnitudes but opposing signs despite the differences in the relative perturbation sizes (see Appendix \ref{appendix:Model Details}).} 	
	Whilst the local SSE [Fig. \ \ref{fig:Wires Perp}b)] represents a net response of all available magnon modes  {with a finite polarization perpendicular to the Pt wire}, the non-local SSE [Fig. \ \ref{fig:Wires Perp}c)] only has contributions from magnon modes that propagate across 500 nm (see also Appendix \ref{appendix:Distance Dependence}) and can be detected. 
	The magnon propagation length in AFMs has been shown to be dependent on the frequency, indicating that the filtering out of higher frequency magnon modes with decreasing temperature  {will correspondingly affect the magnon propagation length at a given temperature}  \cite{Cramer2018}.
	Meanwhile, the application of a field, even below the spin-flop, will significantly alter the frequencies of the contributing modes [Fig. \ \ref{fig:magnon modes}a)]. 
	 {Quantitatively, the application of the field at \SI{200}{\K} initially breaks the degeneracy, enabling spin transport which increases as the degeneracy is broken further. 
	As the frequencies of the magnon modes are changed by the magnetic field, the relative populations will also change. As discussed previously, different magnon modes contribute to the net spin Seebeck with differing signs \cite{Rezende2018, Li2019}, and the propagation lengths will continuously change with changing frequency \cite{Cramer2018}. Eventually the mode character changes at the spin-reorientation and the signal adopts a linear increase with field due to the emerging $\vec{m}$. Hoogeboom \textit{et al.} \cite{Hoogeboom2020} very recently discussed the transport of thermally excited magnons in the easy-plane antiferromagnet NiO, where the magnetic structure is very similar to the configuration we have above $H_{cr}$ or above $T_M$. They found reasonable agreement between the non-local transport and the field dependence of the lowest k=0 magnon mode at low temperatures. However, Hoogeboom \textit{et al.} apply a magnetic field within the antiferromagnetic easy-plane as compared to our field, which is maintain along $\vec{x}$ at all times. This then leads to the magnon dispersions of $\alpha$-Fe$_2$O$_3$ above $H_{cr}$ and NiO being very different. Given this complexity, determining the exact contributions of different modes is out of the scope of this work, which instead focuses on the contributions of the magnetic order parameters to the net transport.
	Qualitatively, we can interpret the experimental results in terms of the population of different magnon modes [Fig.\ \ref{fig:magnon modes}] which contribute to the signal \cite{Rezende2018}. 
	With reducing temperature, the available magnon modes should correspondingly be reduced.
	If the non-local signal observed here is dominated by lower energy magnon modes, a reduction in temperature should not lead to significant changes to the signal other than the critical spin-flop field increasing until we decrease below the thermal energy required to populate the second lowest mode, where the signal should exhibit a peak \cite{Wu2016, Rezende2018}. 
	We show in Fig. \ \ref{fig:Wires Perp}d) and \ref{fig:Wires Perp}e) the non-local signal, normalized to the heating power based on the heater resistance for several temperatures. 
	With decreasing temperature below $T_M$ we observe that the overall shape of a positive increase followed by a sign inversion is a common trend, however, the transition between the two regimes becomes increasingly smooth at lower temperatures. 
	Above $T_M$, where $\alpha$-Fe$_2$O$_3$ adopts a canted easy-plane anisotropy, the application of $\vec{H}$ (along the same direction as below $T_M$) leads to a linearly increasing moment parallel to the field, represented by the linearly increasing NL-SSE signal observed  {(confirming our previous measurements of a linearly increasing NL-SSE above $H_{cr}$ where the antiferromagnetic configurations are similar)}.
	The similar shapes below $T_M$ would indicate that similar magnons are contributing at each temperature, and these magnons can propagate over large distances with no evidence of a distance-dependent sign change \cite{Hoogeboom2021} (see also Appendix \ref{appendix:Distance Dependence})}. 
	The magnitude of the NL-SSE is shown in Fig. \ \ref{fig:Wires Perp}e) with the spin-reorientation field \cite{Lebrun2019} and the field where the signal inverts indicated. 
	As exemplified in Fig. \ \ref{fig:Wires Perp}c), the spin-reorientation and inversion fields are not the same and diverge with decreasing temperature below $T_M$. 
	
	\section{Thermal Magnon Perpendicular to the Easy Axis}\label{sec:transport perp}
\begin{figure*}[ht!]
		\includegraphics[width = 12.1cm]{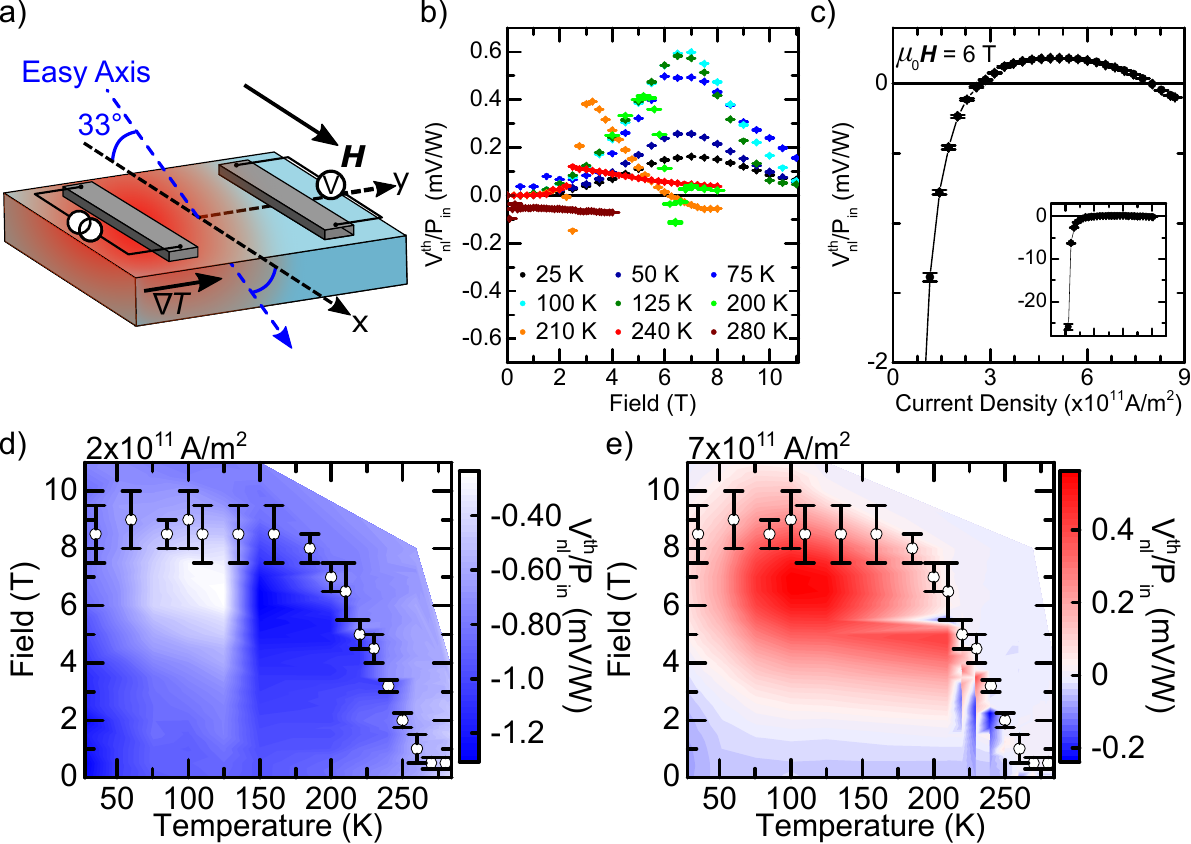}
		\caption{\label{fig:Wires parallel}a)) Non-local schematic for wires parallel to  $\vec{x}$. The magnetic field is parallel to the easy-axis in-plane projection and perpendicular to the thermal gradient $\nabla T$. b) Non-local antiferromagnetic spin Seebeck effect as a function of field for temperatures below $T_M$.   {The magnitude of $\vec{H}$ stated is that along $\vec{x}$.} c) Current dependence of the non-local signal normalized to the input power. The inset shows the same data but on a larger y-axis due to the divergent behavior at low input currents d) - e) Phase diagram of the non-local spin Seebeck signal as a function of field and temperature for two current densities, \SI{2e11}{\A\per\square\m} and \SI{7e11}{\A\per\square\m} respectively. The circles represent the spin-flop field taken from Ref.\ \protect\cite{Lebrun2019}.}
	\end{figure*}

	 {Next}, we show in Fig. \ \ref{fig:Wires parallel}a) a non-local geometry now rotated by \SI{90}{\degree}, such that $\nabla T$ and the transport direction  $\vec{y}$ are perpendicular to both $\vec{H}$ and the in-plane easy-axis component, which is parallel to  $\vec{x}$.
	 {For this geometry, the perpendicular alignment of the magnetic field and the transport direction means that a magnetization will not appear directed along $\vec{y}$ for any magnitude of applied field, allowing us to completely exclude any significant contributions.}
	The absence of $\vec{m}$ is confirmed through measurements of the local SSE (see Appendix \ \ref{appendix:local wires para}), where there is no discernable signal increasing with magnetic field, consistent with previous reports for the AFM-SSE and expected from the parallel alignment of $\vec{H}$ and the voltage detection \cite{Wu2016, Seki2015}. 
	We can also confirm this numerically by calculating in Appendix\ \ref{appendix:Model Details} the different components of the magnetization and \Neel{} vector under an applied magnetic field. 
	The geometry can also be investigated in the non-local configuration where we expect no significant transport to occur perpendicular to the magnetic anisotropy axis as previously discussed \cite{Hoogeboom2020, Muduli2021, Lebrun2018, Rezende2018, Bender2017}. 
	However, as seen in Fig. \ \ref{fig:Wires parallel}b), we see for all temperatures a significant signal above and below $T_M$  {with a drastically different evolution to the previous geometry}. 
	 {At temperatures closer to the Morin transition -- such as for example \SI{210}{\kelvin} -- the NL-SSE is initially constant around zero at low magnetic fields. We observe that, as we approach $H_{cr}$, the NL-SSE shows a sharp negative dip before reaching a smooth maximum at around \SI{3}{\tesla} (close to the measured value of $H_{cr}$ at the temperature for a \SI{33}{\degree} angle between $\vec{H}$ and the easy-axis \cite{Lebrun2019}). Following this peak, the signal smoothly and gradually decreases with increasing magnetic field. We suggest that the sharp oscillations of the signal at fields below $H_{cr}$ come from fluctuations of the N\'eel vector as the magnetic anisotropy is compensated by the applied magnetic field. Following the spin-flop, where the N\'eel vector is now aligned parallel to the transport direction (see Appendix\ \ref{appendix:Model Details}), the increasing magnetic field will lead to a canting of the magnetic sublattices in the direction of $\vec{H}$. This canting will serve to reduce the projection of $\vec{n}$ on the transport direction, similarly reducing the observed signal. We can also exclude to projection of a magnetization along $\vec{y}$ for any applied field along $\vec{x}$ (See Fig.\ \ref{fig:equilibriumPosition_n and m} in the Appendix). However, as we move to temperatures away from $T_M$, we no longer observe the fluctuation of the signal before the peak.}
	Instead, $V_{th}^{nl}$ continues to increase from \SI{0}{\tesla} and peaks at some magnetic field value,  {where the peak values moves to larger magnetic fields with decreasing temperature}, indicating that this peak is related to the magnetic structure of the antiferromagnet and thus, intrinsically the antiferromagnetic anisotropy \cite{Lebrun2019, Morrison1973}. 
	 {We note that, whilst the peak closer to $T_M$ coincides with the value of $H_{cr}$, the peak at lower temperatures occurs at lower magnetic fields than measured for $H_{cr}$ (e.g. \SI{7}{\tesla} at \SI{100}{\kelvin} while $H_{cr}$ is closer to \SI{9}{\tesla}) \cite{Lebrun2019}. This is an effect of the finite angle between $\vec{H}$ and the easy-axis leading to a continuous rotation of $\vec{n}$ across a larger field range \cite{Lebrun2019}. If we reduce the angle between $\vec{H}$ and the easy-axis, we resolve a sharp transition at $H_{cr}$ (Appendix\ \ref{appendix:Beta Plane}).}
	The decrease at high magnetic fields, may also be related to the magnon gap appearing above the spin-flop, similar to the high-field suppression of the SSE seen in ferrimagnetic YIG \cite{Cornelissen2016}.
	However, this behavior depends on the magnitude of the input power, where we see that our signal changes sign with increasing power, which we show in Fig. \ \ref{fig:Wires parallel}c) at \SI{200}{\K} and an applied field equal to the spin-reorientation field ($\mu_0$$\vec{H} = \SI{6}{\tesla}$) \cite{Lebrun2019, Lebrun2018}. 
	A similar effect has been reported for ferrimagnetic YIG, characterizing a transition between a more localized and non-local SSE \cite{Cornelissen2015}, however this is very different to what we observed for the orientation of Fig. \ \ref{fig:Wires Perp} (see Appendix \ref{appendix:Current Dependence}).
	Furthermore, we see that at very large heating power, we again find a negative signal. 
	Without normalizing to $P_{in} = R_{SMR}I^2$, we observe a predominantly quadratic dependence of the voltage, confirming that the observed signal is indeed thermal in origin. 
	This behavior is present across all temperatures investigated. 
	 {It was stated by Shan \textit{et al.} \cite{Shan2016} that the non-local SSE depends on the boundary conditions of the magnon current, meanwhile recent work has also highlighted the impact of the antiferromagnetic domain structure on magnon transport \cite{Hoogeboom2020, Ross2020}. The presence of such domain walls could then alter the boundary conditions of the propagating magnon currents. However, we can exclude the possibility of antiferromagnetic domain walls between our Pt wires given that we utilize a bulk $\alpha$-Fe$_2$O$_3$ crystal with domains of hundreds of \si{\micro\m} \cite{Lebrun2018}. Furthermore, the strong temperature dependence of the anisotropy of $\alpha$-Fe$_2$O$_3$ \cite{Lebrun2019, Morrison1973} could also lead to temperature induced changes of the antiferromagnetic state. This could possibly lead to these sign changes as a function of the input power through altering the magnon chemical potential and locally changing the anisotropy, especially close to $T_M$ \cite{Flebus2019}.}
	We summarize the field and temperature dependence in Fig. \ \ref{fig:Wires parallel}d) and \ref{fig:Wires parallel}e) alongside the spin-reorientation field for two input current densities. \\
	%
	%
	%
	%
		
	Given that the magnetic field direction and wire separation are maintained, the available and excited magnon modes (Fig. \ \ref{fig:magnon modes}) should be identical to the geometry of Fig. \ \ref{fig:Wires Perp}a). 
	This then raises the question of what leads to the net spin transport perpendicular to the applied magnetic field. 
	We find that we cannot reproduce these experimental findings using our theoretical model (see next section) so it is clear that the origin of thermally excited magnon transport perpendicular to the easy-axis requires further study. 
	It may be that the differences lie in magnon modes further in the Brillouin zone than just the $k=0$ modes.
	Experimentally, we are detecting a voltage that encompasses all the available magnon modes propagating perpendicular to the easy-axis  {with a finite component of the polarization perpendicular to the Pt detector.} 
	It has been shown for ferrimagnetic YIG that the local SSE has a significant contribution not only from the $k=0$ mode, but also from magnons throughout the Brillouin zone \cite{Rezende2018, Kehlberger2015}.
	Furthermore, there are asymmetries in the spin-wave dispersion in $\alpha$-Fe$_2$O$_3$ between spin-waves parallel and perpendicular to the easy-axis \cite{Samuelsen1970}.
	Measurements of the spin-wave dispersion in $\alpha$-Fe$_2$O$_3$ both parallel and perpendicular to the easy-axis direction find velocities of 24 and 31 km/s, respectively, as well as a gap at the edge of the Brillouin zone between the acoustic and optical branches \cite{Samuelsen1970, Samuelsen1970_cr2o3}. 
	The spin-waves are at far higher energies than those of the otherwise similar antiferromagnet Cr$_2$O$_3$ \cite{Samuelsen1970, Samuelsen1970_cr2o3}, which has been the subject of both local \cite{Seki2015, Muduli2021} and non-local \cite{Yuan2018, Muduli2021} SSE measurements. 
	Evidence of spin transport perpendicular to the easy-axis in this case may also be related to the exchange interaction of $\alpha$-Fe$_2$O$_3$ as compared to for instance Cr$_2$O$_3$. 
	The strongest exchange paths of $\alpha$-Fe$_2$O$_3$ are the weakest in Cr$_2$O$_3$ (in other words, the relative strength of the exchange interaction for the nearest neighbors that significantly contribute giving rise to different exchange constants $J_i$ where $i$ is the order of neighbor) due to differences in the cation-cation distance of these two antiferromagnets\cite{Samuelsen1970, Samuelsen1970_cr2o3}. 
	The exchange paths are also different to the previously studied fluorides MnF$_2$ \cite{Samuelsen1970, Wu2016} and FeF$_2$ \cite{Samuelsen1970, Li2019} so may contribute to the observed differences. 
	Finally, there is the possible contributions arising from the additional DMI present in the system, an interaction absent in all previously investigated antiferromagnets. 

	\section{Atomistic spin model}  \label{sec:ASM}
	 {Having established that significant magnon transport is possible in antiferromagnetic $\alpha$-Fe$_2$O$_3$ mediated by the N\'eel vector both parallel and perpendicular to the easy-axis, we develop in this section a toy model for the transport in this material.}
	For this, we utilize the following classical, atomistic spin model \cite{Nowak07_SpinModels}:
	it comprises $N = N_x \times N_y \times N_z$ Heisenberg spins, i.e.\ normalized magnetic moments $\vec{S}^l = \vec{\mu}^l/\muS$, which can point in every three-dimensional spatial direction.
	These spins are positioned on regular lattice sites $\vec{r}^l$, where we assume a simple cubic (sc) lattice with lattice constant $a$.
	The spin Hamiltonian reads
	\begin{widetext}
		\begin{align}
			 {\mathcal{H}} =   - \frac{1}{2} \sum_{n=1}^N \sum_{ \substack{m \in \\ \operatorname{NB}(n)}} \left(\vec{S}^n\right)^\dagger \bm{\mathfrak{J}}^{nm} \vec{S}^m 
			- \sum_{n=1}^N \left[ d_z^{(2)}\left(S_z^n\right)^2 + d_z^{(4)}\left(S_z^n\right)^4 \right] 
			- \muS\sum_{n=1}^N\vec{S}^n\cdot\vec{B},
		\end{align}
	\end{widetext}
	taking into account the exchange interaction, quantified by the exchange Matrix $\bm{\mathfrak{J}}^{nm}$, where the sum over $m$ runs over the $N_\mathrm{nb}$ neighbors of $n$, labelled by $\mathrm{NB}(n)$,---counting each pair interaction twice in the double sum.
	These matrices split into the isotropic Heisenberg contribution (scalar part), the two-ion anisotropy (traceless, symmetric part) and the Dzyaloshinskii-Moriya interaction (antisymmetric part).
	Furthermore, a uniaxial anisotropy with respect to the $z$ direction with anisotropy constants $d_z^{(2)}$ and $d_z^{(4)}$ is included, parametrizing second- and fourth order contribution.
	A homogeneous magnetic field $\vec{B}$ enters the Zeeman term. \\
	
	The Landau-Lifshitz-Gilbert (LLG) equation of motion \cite{Landau35_LL_equation,Gilbert55_Gilbert_damp,Gilbert04_Gilbert_damp_IEEE} governs the time evolution:
	\begin{equation}
		\begin{gathered}
			\frac{\mathrm{d} S^l}{\dd t} = -\frac{\gamma}{\mu_S(1 + \alpha^2)} \left[ S^l \times \left( H_\mathrm{eff}^l + \alpha S^l \times H_\mathrm{eff}^l \right) \right] , \label{eq:LLG} \\
			H_\mathrm{eff}^l = - \frac{\partial \mathcal{H}}{\partial S^l} + \xi^l
		\end{gathered}
	\end{equation}
	describing the motion of a spin in its effective field $\vec{H}_\mathrm{eff}^l$, where $\gamma$ is the gyromagnetic ratio and $\alpha$ the Gilbert damping constant.
	Finite temperatures enter the model via a Gaussian white noise $\vec{\xi}^l$ \cite{Brown63_ThermalFluctuactionsMagnParticles}, satisfying
	\begin{gather*}
		\left\langle \vec{\xi}^l(t) \right\rangle = 0 \mkern9mu \text{and} \\ 
		\left\langle \xi^l_\beta(t)\xi^k_\eta(t') \right\rangle = \frac{2\muS\alpha\kBT_l}{\gamma} \delta_{lk}\delta_{\beta\eta}\delta(t - t'), \\ \beta,\eta\in\{x,y,z\} .  \nonumber
	\end{gather*}
	Numerically the LLG equation is solved by the stochastic Heun method \cite{Nowak07_SpinModels}.\\
	
\begin{figure}[ht!]
		\includegraphics[width = 8cm]{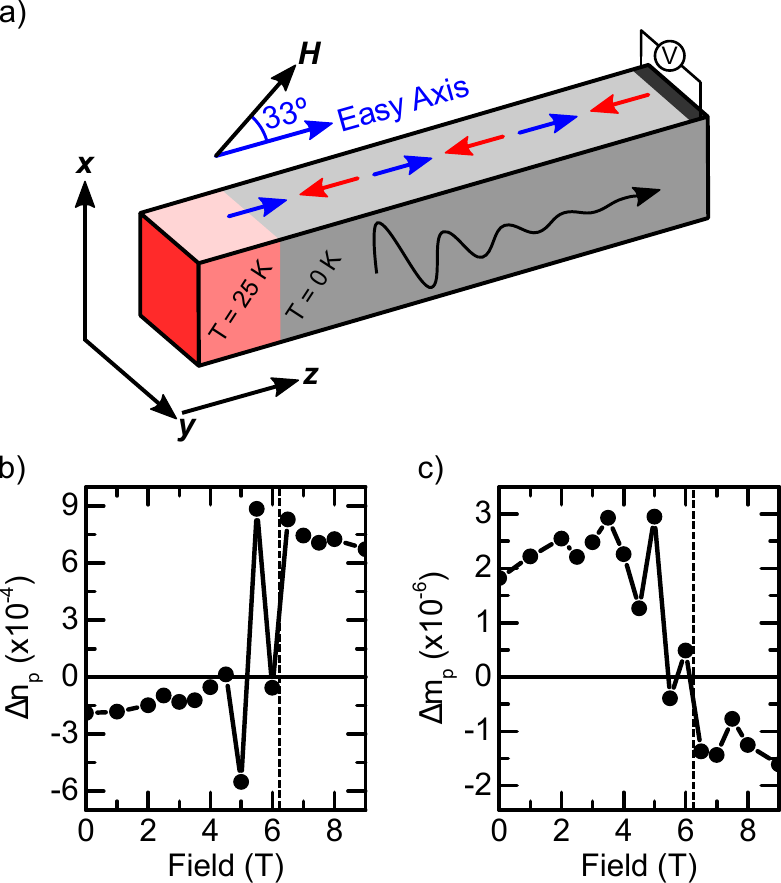}
		\caption{\label{fig:theory figure}a)) Geometry used for calculations with a size $N_x \times N_y \times N_z = \num{16} \times \num{16} \times \num{12288}$. A temperature step of \SI{25}{\K} was used to excite the magnetic ordering and we probe the perturbation from equilibrium of the \Neel{} vector and magnetization some distance from this excitation. The easy-axis is aligned along the transport direction and a magnetic field $\vec{H}$ is applied at \SI{33}{\degree} to the easy-axis. b) Perturbation of the \Neel{} vector from equilibrium ($\vec{n}_0$) as a function of the applied field, projected along the magnetic field, where $\Delta n_\mathrm{p}$ is defined in the main text. b) Perturbation of the magnetization from equilibrium ($\vec{m}_0$) as a function of the applied field, projected along the magnetic field, where $\Delta m_\mathrm{p}$ is defined in the main text.  {The dashed lines indicates the spin-flop field of the system.}}
	\end{figure}

	The four atoms---denoted A, B, C, D---of the $\alpha$-Fe$_2$O$_3$ unit cell are placed in the form of layers onto the sc lattice, i.e.\ if $(x,y,z)$ are the lattice indices, atoms of sublattice A are positioned on sites with $z \bmod 4 = 0$, those of sublattice B on sites with $z \bmod 4 = 1$ and so on.
	The result is an antiferromagnet layered along the $z$ direction.
	Note that as in the actual hematite unit cell, sublattices A and D align parallel, and these are antiparallel to B and C \cite{Morrish94_CantedAFM}.
	Our system is of a size $N_x \times N_y \times N_z = \num{16} \times \num{16} \times \num{12288}$, when orientated to study the transport along the easy-axis [Fig. \ \ref{fig:theory figure}a)]. 
	The finite cross section in $x$- and $y$ direction ensures thermal stability of the magnetic order at the temperatures we investigate in this study.\\
	%
	%
	%
	%
	
	We parameterize this model as discussed in the Appendix \ref{appendix:Model Details}, where we also calculate the $x,y,z$ components of both $\vec{n}$ and $\vec{m}$ under an applied magnetic field.	
	The thermal excitation is modelled by a temperature step similar to previous works \cite{Ritzmann14_MagnonPropagation,Ritzmann17_SSE_TwoSublMagnets}
	\begin{align}
		T_l = \left\{ \begin{array}{ll}
			\SI{25}{\K} & \text{for } z^l < 614a \\
			\SI{0}{\K}  & \text{else}
		\end{array} \right.  .
	\end{align}
	We also tested a linear temperature gradient, which delivers qualitatively and quantitatively the same results, such that we use a temperature step in the following because of its simplicity.
	The thermal fluctuations in the hot area excite magnons, which propagate into the cool area---along the $z$ direction---and are damped away due to Gilbert damping.
	The geometry is sketched in Fig. \ \ref{fig:theory figure}a).\\
	
	Each sublattice carries a normalized magnetization $\vec{m}^X$, $X\in\{\mathrm{A},\mathrm{B},\mathrm{C},\mathrm{D}\}$, from which the system's \Neel{} vector $\vec{n} = \frac{1}{4}\left( \vec{m}^\mathrm{A} - \vec{m}^\mathrm{B} - \vec{m}^\mathrm{C} + \vec{m}^\mathrm{D} \right)$ and the magnetization $\vec{m} = \frac{1}{4}\left( \vec{m}^\mathrm{A} + \vec{m}^\mathrm{B} + \vec{m}^\mathrm{C} + \vec{m}^\mathrm{D} \right)$ follow.
	The thermal excitation in steady state leads---on average---to a change of both properties as a function of position $z$.
	We study this change as a function of magnetic field:
	$\vec{n}_0$ and $\vec{m}_0$ denote the equilibrium values for a given magnetic field, and we define the deviation of the \Neel{} order- and spin accumulation, by
	\begin{align*}
		\Delta \vec{n}(z) & = \left\langle \vec{n}(z) \right\rangle - \vec{n}_0  \\
		\Delta \vec{m}(z) & = \left\langle \vec{m}(z) \right\rangle - \vec{m}_0 .
	\end{align*}
	The static magnetic field is tilted by $\beta = \SI{33}{\degree}$ with respect to the easy-axis: $\vec{B} = B_\mathrm{s} \cdot (\sin(\beta), 0, \cos(\beta))$ [as shown in Fig. \ \ref{fig:theory figure}a))].
	This same angle between the easy-axis transport direction and the field allows us to compare the calculations with the experiment: in the model the transport is along the easy-axis and the magnetic field is tilted, whereas in the experiment this is reversed.
	This difference is due to the requirements of the numerical treatment.
	However, since the detected signal in the experiment is most likely dominated by the components along the transport direction, i.e.\ the components along the field, we can thus assume that the projection of the accumulations onto the direction of the magnetic field direction is the property to evaluate for a comparison  of model and experimental results:
	\begin{align*}
		\Delta n_\mathrm{p} = \Delta \vec{n} \cdot \vec{e}_{B} \quad\text{and}\quad
		\Delta m_\mathrm{p} = \Delta \vec{m} \cdot \vec{e}_{B} .
	\end{align*}\\

	Fig. \ \ref{fig:theory figure} depicts the resulting \Neel{} order- [Fig. \ \ref{fig:theory figure}b)] and spin-accumulation [Fig. \ \ref{fig:theory figure}c)] at position $z = 2000a$ as a function of magnetic field.
	 {Around the spin-flop ($B_{sf} \approx \SI{6.4}{T}$ in the model), we see a large scattering of both $\Delta n_p$ and $\Delta m_p$, however, this is due to numerical uncertainties, which stem from the fact that here it is very hard to calculate the respective equilibrium positions of N\'eel vector and magnetization.
		At the same time this also makes it hard to ascertain proper error bars in this region.
		Looking at the general trends away from the spin-flop field, both show a non-monotonic behavior and a sign change, which is, however, always in the vicinity of the spin-flop field.
		Furthermore, both start at a non-zero value for zero field, which is expected for the N\'eel accumulation, but not the spin accumulation.
		This offset originates from numerical inaccuracies in the parameters being used -- specifically in the exchange matrices.
		These have small but finite imbalances among the different sublattices, which cause a very small spurious magnetization at finite temperature as the sublattices do not demagnetize uniformly with increasing temperature.
		These numerical errors only occur at finite temperatures and are therefore not evident in the equilibrium magnetization curve (Fig.\ \ref{fig:equilibriumPosition_n and m} in the Appendix).
		But they are visible when compared to the small non-equilibrium spin accumulation due to the temperature gradient.
		More accurate exchange matrices would eliminate this offset, which is however not significant for our findings.
		Above the spin-flop, the spin accumulation seems to increase with field, whereas the N\'eel accumulation tends to decrease, however, in either case the field dependence is rather weak here.
		It is worth noting the $\Delta n_p$ is several orders of magnitude larger than $\Delta m_p$.
		This is due to the large exchange energy present in antiferromagnets and the relation between the N\'eel vector and the magnetization.
		A perturbation of the N\'eel vector is related to the magnetic anisotropy along the easy-axis alongside the exchange energy, while perturbations of the magnetization are competing with the strong exchange energy.
		Therefore it is easier to alter the N\'eel vector by an external bias (as a temperature gradient for instance) compared to the magnetization.}\\
	
	We next qualitatively compare the experimental results we obtain, with several key features of an increase and maximum of the signal below the spin-flop, an inversion and a linear increase above $H_{cr}$ with the theoretical calculations of the \Neel{} order- and spin- accumulations in Sec.\ \ref{sec:ASM} [Fig.\ \ref{fig:theory figure}b) -- c)].
	We find agreement between the atomistic results and experiments for some features, but also differences:
	 {The non-monotonic progression of the experimental data allows for a maximum of the detected signal at a finite field away from the spin-flop field. Meanwhile in our toy model, if we assume that below the spin-flop the spin accumulation dominates the signal, the spin-accumulation increases slightly up to the vicinity of the spin-flop field.
	Furthermore, as in the experiments, there is a sign change in both $\Delta n_p$ and $\Delta m_p$, but not significantly below the spin-flop field. }
	A monotonic increase of the signal observed in the experiment at high fields is also consistent at least with the spin accumulation [Fig. \ref{fig:theory figure}c)], but not in line with the \Neel{} accumulation [Fig. \ref{fig:theory figure}b)],  {highlighting the importance of the net magnetization above the spin-flop}.
	Measurements of the thermal magnon transport performed for varying angles between $\vec{H}$ and the easy-axis show a varying behavior which may hint at the origin of the transport but is outside the scope of the current work [see Appendix \ \ref{appendix:Beta Plane}].\\
		
	While key features are reproduced, there are many reasons for the differences between experiments and the model:
	In the theory we have a clean magnetic order, i.e.\ domain formation can be ruled out \cite{Ross2020, Hoogeboom2021},  {however, considering that we make use of a bulk crystal here, any domains are far larger than the transport scales we are considering \cite{Lebrun2018}}. 
	Also we do not account for any disturbances that might be present at the hematite/Pt interface, i.e.\ roughness or proximity effects.
	The calculation is furthermore carried out at a base temperature of zero Kelvin, which arguably also makes a difference.
	Last but not least there is a large uncertainty, how each of the accumulations contributes to the experimentally detected signal.
	The importance of each contribution may also vary with field as the \Neel{} vector reorients by \SI{90}{\degree} at the spin-flop.	
	Hence, the complex field and temperature dependence and also differences between experiment and model underline the complex behavior of the spin transport in this system, which requires a deeper understanding of the microscopic details.
	
	\section{Conclusion}\label{sec:conclusion}	
	In conclusion, we have observed the transport of thermally excited magnon spin currents in the absence of a significant field induced magnetization parallel to the transport direction. 
	The contribution to the net signal of different magnon modes leads to a complex field dependence below the spin-flop field. 
	Above the spin-flop field, a signal with an amplitude proportional to the field induced magnetization appears. 
	We also observe transport perpendicular to the applied magnetic field, where no magnetization appears, contrary to previous reports on uniaxial antiferromagnets. 
	Our results highlight the importance of the \Neel{} order spin Seebeck conductance for the complex behavior of magnon transport in antiferromagnetic materials under an applied magnetic field offering additional routes for tuning the transport of magnonic spin currents in antiferromagnetic insulators.

	%
	%
	%
	%
	%
	%
	%
	%
	%
	\appendix
	\section{Parameter Choice for Atomistic Spin Model}\label{appendix:Model Details}
	Finding suitable parameters for the atomistic calculations is challenging: here we use a combination of ab-initio based parameters and fits to experimentally known properties.
	First, $\gamma = \gamma_e$ and $\muS = 4.9\mu_\mathrm{B}$ with $\gamma_e$ the absolute value of the free-electron gyromagnetic ratio and $\mu_\mathrm{B}$ the Bohr magneton follow from experiments.
	Second, the exchange matrices for the real hematite structure are calculated by density functional theory (DFT) in terms of the screened Korringa--Kohn--Rostoker method \cite{Zabloudil2005, Deak2011,Szunyogh2011}. 
	These interactions turn out to be rather long ranged and since an accurate description of hematite is beyond the scope of work, we instead have to map these to our toy-model hematite.
	This is carried out by first mapping the DFT values to mean-field parameters (one matrix for each interaction of one sublattice with another), then this is partitioned on the existing interactions in the toy model.
	This partitioning is done such that it roughly represents the relative importance of the interactions strength with distance of the original DFT results, e.g.\ the isotropic part is strongest for next-nearest neighbours.\\
	
	The on-site anisotropy constants are chosen according to $\nicefrac{d_z^{(2)}}{d_z^{(4)}} = 4$, a ratio known from mean-field modelling in the literature \cite{Morrish94_CantedAFM}, and such that the spin-flop fields are close to the experimental values \cite{Morrison1973,Lebrun2019}.\\
	
	For the Gilbert damping we assume $\alpha = \num{5e-4}$, which is likely higher than the actual value of a high-quality single crystal \cite{Lebrun2020}, however, we choose a value that does not require excessively large systems for the numerical studies.\\
	
\begin{figure}[ht!]
		\includegraphics[width = 8cm]{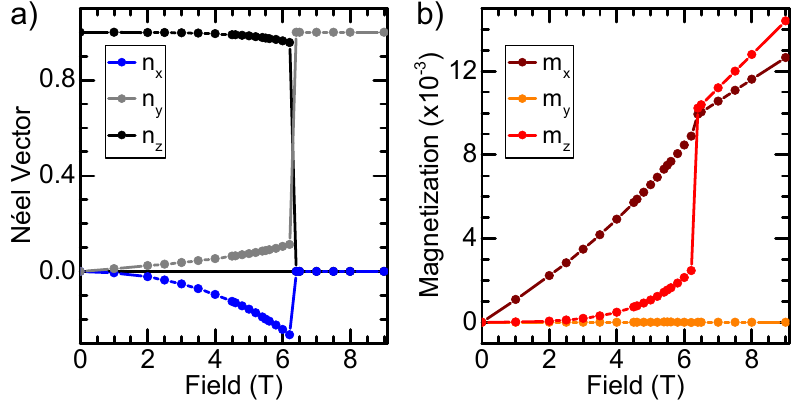}
		\caption{\label{fig:equilibriumPosition_n and m}a) $x$, $y$, and $z$ components of the \Neel{} vector under an applied magnetic field at \SI{33}{\degree} to the easy-axis for our choice of parameters. We find the spin-flop to occur at $\mu_0\vec{H}= \SI{6.4}{\tesla}$, similar to the experimental observation. b) $x$, $y$, and $z$ components of the magnetization under an applied magnetic field for our choice of parameters. Above the spin-flop field, a large component of $\vec{m}$ appears parallel to the field. }
	\end{figure}

	Having chosen an appropriate parameter set, we can calculate the equilibrium position of the \Neel{} vector and magnetization under an applied field, devolving both into their respective $x$, $y$ and $z$ components, Fig. \ \ref{fig:equilibriumPosition_n and m}. 
	%
	%
	%
	%
	
	\section{Current Dependence for Transport Parallel to the Easy Axis}\label{appendix:Current Dependence}
	In Fig. \ \ref{fig:Wires parallel}, a significant difference was observed between two current densities passed through the Pt injector wire, with a change of sign at low and high current densities. 
	In Fig. \ \ref{fig:Wires Perp, Current Dependence} we show the local and non-local spin Seebeck signal normalized to the input heating power for the geometry shown in Fig. \ \ref{fig:Wires Perp}a) for two current densities at \SI{200}{\K}. 
	We observe that the trends of both are identical for the two current densities, unlike for the transport perpendicular to the easy-axis. 
	The absolute magnitude of both, when normalized for the heating power, are also identical within the error bars confirming the thermal origins of the measured signals with no hint of a sign change. 	
	%
	%
	\begin{figure}[ht!]
		\includegraphics[width = 8cm]{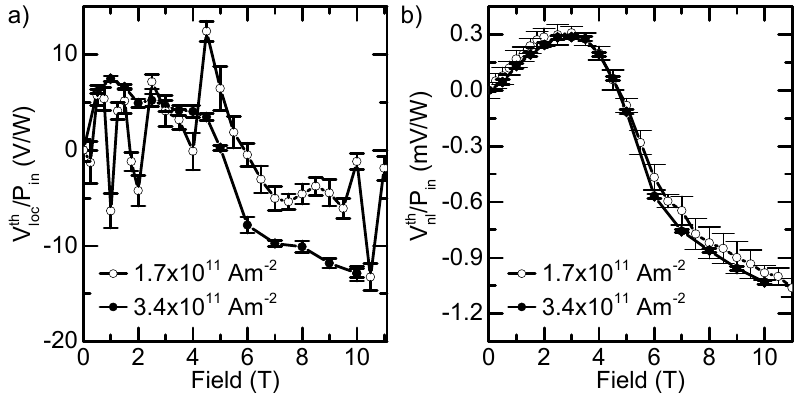}
		\caption{\label{fig:Wires Perp, Current Dependence}a) Local spin Seebeck effect measured for the geometry in Fig. \ \ref{fig:Wires Perp}a) as a function of the applied magnetic field for two current densities. b) Non-local spin Seebeck effect measured for the geometry in Fig. \ \ref{fig:Wires Perp}a) as a function of the applied magnetic field for two current densities. Error bars represent the standard deviation of the data point.}
	\end{figure}
	%
	%
	
	\section{Persistence of the Non-local Spin Seebeck Signal for Transport Parallel to the Easy Axis}\label{appendix:Distance Dependence}
	In addition to the wire separation of \SI{500}{\nm} discussed in the main text, we also investigate the non-local spin Seebeck signal for wires separated by \SI{7}{\micro\m}. 
	By increasing the separation to this, we can reduce the possibility of detecting a more ``localized'' non-local signal due to thermal gradients under the detector wire. 
	We show the NL-SSE signal in Fig. \ \ref{fig:Wires Perp, Distance Dependence} for \SI{500}{\nm}, replicating the data shown in Fig. \ \ref{fig:Wires Perp}d), and for \SI{7}{\micro\m} across a large range of temperatures. 
	We observe that the effect of a magnetic field on the NL-SSE is the same for both wire separations emphasizing that the signal that we detect is due to the magnetic and transport properties of the hematite below. 
	The reduction in the signal and increase in the noise between the two wire separations further supports that the signal originates from the diffusive transport of magnons at all temperatures, with no evidence of non-diffusive transport as has been previously suggested for thermal magnons \cite{Yuan2018}. 
	%
	%
	\begin{figure}[ht!]
		\includegraphics[width = 8cm]{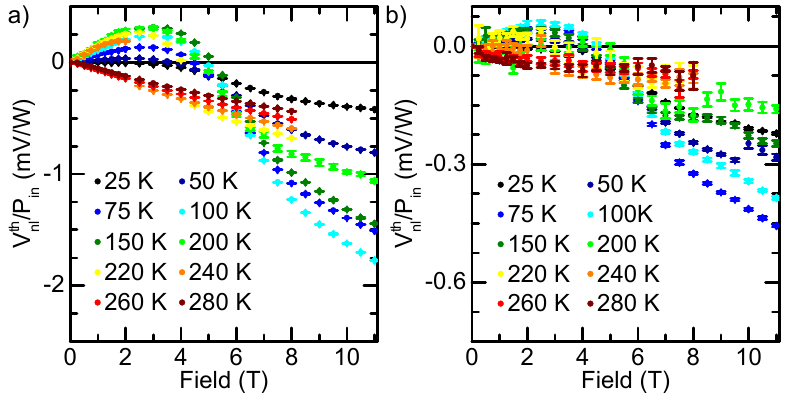}
		\caption{\label{fig:Wires Perp, Distance Dependence}Non-local spin Seebeck signal parallel to the in-plane projection of the easy-axis in hematite for a magnetic field applied parallel to the transport direction across a distance of a) \SI{500}{\nm} or b) \SI{7}{\micro\m}.   {The magnitude of $\vec{H}$ stated is that along $\vec{x}$.} The general shape and trends are identical between the two distances. The error bars represent the standard deviation of the data point.}
	\end{figure}
	%
	%
	
	\section{Rotation of a Magnetic Field Out of the Sample Plane}\label{appendix:Beta Plane}
	In the main text, we discuss the thermal transport parallel to the in-plane projection of the easy-axis (Fig. \ \ref{fig:Wires Perp}) and perpendicular to the easy-axis (Fig. \ \ref{fig:Wires parallel}) under an applied magnetic field. 
	However, we maintain the magnetic field parallel (perpendicular) to the transport direction and such, our measurements in the main text take place with a finite angle of \SI{33}{\degree} between the field $\vec{H}$ and the \Neel{} vector $\vec{n}$. 
	We can probe the non-local spin Seebeck signal not only for this configuration of $\vec{H}$ and $\vec{n}$, but instead rotate $\vec{H}$ through an angle $\beta$ out of the sample plane (Fig. \ \ref{fig:Wires Perp, Beta Plane}a)), where $\beta = 0$ occurs for $\vec{H}$ within the sample plane parallel to  $\vec{x}$. 
	We show in Fig. \ \ref{fig:Wires Perp, Beta Plane}b) and Fig. \ \ref{fig:Wires Perp, Beta Plane}c) the local SSE (red) and NL-SSE for two wire separations (see Appendix \ref{appendix:Distance Dependence}) for a fixed magnetic field of \SI{2}{\tesla} and \SI{8}{\tesla} respectively. 
	Alongside the experimental data points, we fit with a $\sin\beta + \beta_0$ function, where the angular symmetry comes from the detection mechanism relying on the inverse spin Hall effect \cite{Lebrun2018, Sinova2015}. \\
	
	%
	%
	\begin{figure}[ht!]
		\includegraphics[width = 8cm]{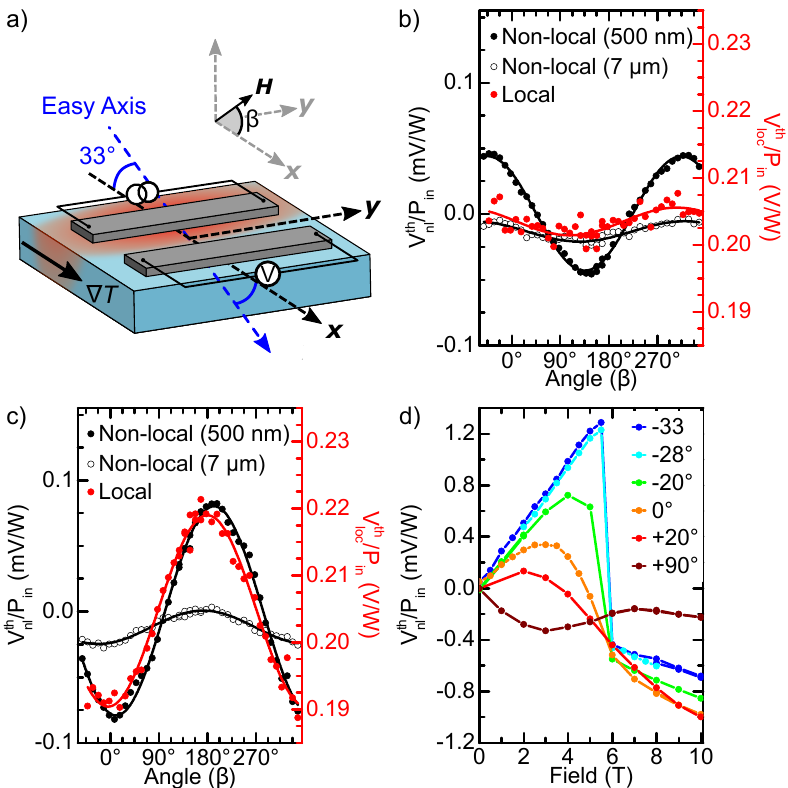}
		\caption{\label{fig:Wires Perp, Beta Plane}a) Non-local geometry for investigating the non-local spin Seebeck effect in hematite for transport parallel to the in-plane projection of the easy-axis. The magnetic field is applied at an angle $\beta$ as indicated where $\beta = \SI{0}{\degree}$ occurs for $\vec{H}$ parallel to  $\vec{x}$. b) Local (red, right axis) and non-local (black closed and open symbols, left axis) spin Seebeck signals at \SI{200}{\K} under a magnetic field of $\mu_0\vec{H} = \SI{2}{\tesla}$ rotated through an angle $\beta$. The solid lines represent fits to a $\sin\beta$ function. c) Local (red, right axis) and non-local (black closed and open symbols, left axis) spin Seebeck signals at \SI{200}{\K} under a magnetic field of $\mu_0\vec{H} = \SI{8}{\tesla}$ rotated through an angle $\beta$. The solid lines represent fits to a $\sin\beta$ function. d) Non-local spin Seebeck signal for a fixed value of $\beta$ for a wire separation of \SI{500}{\nm} at \SI{200}{\K}. }
	\end{figure}
	%
	%
	
	A magnetic field of $\mu_0\vec{H} = \SI{2}{\tesla}$ is below the spin-flop field at this temperature \cite{Lebrun2019, Morrison1973}. 
	There is then correspondingly no significant field induced magnetization due to the low magnetic field and we observe no local spin Seebeck signal. 
	However, we observe a clear non-local signal that follows the expected $\sin\beta + \beta_0$ signal from the spin Seebeck effect, where we find a maximum for $\beta = \SI{-33}{\degree}$. 
	Looking above the spin-flop field (but below the Dzyaloshinskii-Moriya induced spin reorientation) \cite{Morrison1973, Lebrun2019} at $\mu_0\vec{H} = \SI{8}{\tesla}$, a significant local SSE now appears but with the minimum (maximum) closer to the magnetic field lying in-plane than parallel to the easy-axis itself.
	This maximum comes from the field induced magnetization appearing parallel to $\vec{H}$. 
	Looking at the non-local signal though, there is a phase shift between the local and non-local signals, with the minimum (maximum) in the non-local signal occurring for $\sim \SI{10}{\degree}$ ($\sim \SI{190}{\degree}$.
	Instead of fixing the magnitude of $\vec{H}$, we can fix the value of $\beta$ and probe the NL-SSE as a function of the magnetic field which we show in Fig. \ \ref{fig:Wires Perp, Beta Plane}d).
	The most significant signal appears for a magnetic field parallel to the easy-axis at $\beta = \SI{-33}{\degree}$. 
	We observe a linear increase of the observed signal with a sharp, sudden change of sign at the spin-flop field. 
	The transition becomes smoother with increasing angle, with the shape of the signal even changing sign when the magnetic field is applied completely out of the sample plane. 
	With an out-of-plane component of $\vec{H}$, we note that there may also be contributions from other thermo-electric effects such as the Righi-Leduc effect \cite{Thiery2018}, but we would anticipate these to appear with a linear dependence on the applied magnetic field. \\
	
	 {We can additionally probe the non-local spin Seebeck for the geometry discussed in Sec.\ \ref{sec:transport perp} for a magnetic field applied at different angles in the plane perpendicular to $\vec{y}$. In the main text, we observe a significant transport signal perpendicular to the antiferromagnetic easy-axis. Unlike the geometry above, there is no component of the magnetization parallel to the transport direction for any magnitude of the applied field [see Fig.\ \ref{fig:equilibriumPosition_n and m}]. We show in Fig.\ \ref{fig:Wires Para, Beta Plane}b) the NL-SSE for different angles between $\vec{H}$ and the easy-axis. We resolve that when the angle between the two is close to \SI{0}{\degree}, we observe an increasing signal from \SI{0}{\tesla} until the spin-flop field, where a sharp transition results in a small, finite but field-independent voltage. Given the absence of a magnetization, this signal results purely from the transport of thermally excited magnons by the antiferromagnetic N\'eel vector. }

	%
	%
	\begin{figure}[ht!]
		\includegraphics[width = 8cm]{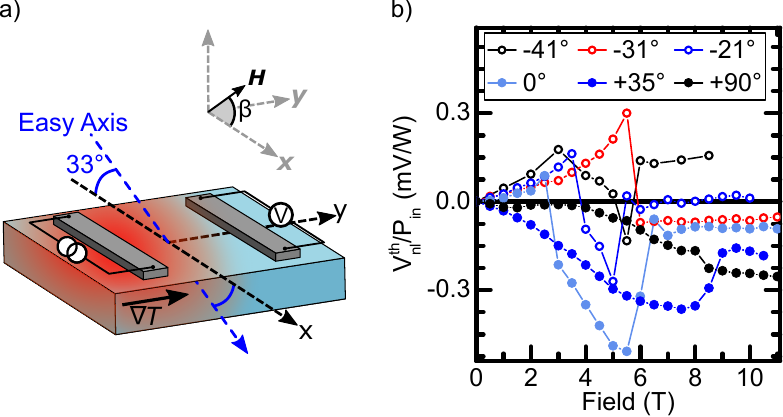}
		\caption{\label{fig:Wires Para, Beta Plane} {a) Non-local geometry for investigating the non-local spin Seebeck effect in hematite for transport perpendicular to the in-plane projection of the easy-axis. The magnetic field is applied at an angle $\beta$ as indicated where $\beta = \SI{0}{\degree}$ occurs for $\vec{H}$ parallel to  $\vec{x}$. b) Non-local spin Seebeck signal for a fixed value of $\beta$ for a wire separation of \SI{500}{\nm} at \SI{200}{\K} and a current density of \SI{2e11}{\A\per\square\m}.}} 
	\end{figure}
	%
	%

	\section{Local spin Seebeck signal for wires parallel to the easy-axis}\label{appendix:local wires para}
	We record the local spin Seebeck signal for the geometry discussed in Fig.\ \ref{fig:Wires parallel}, which we show in Fig. \ \ref{fig:Wires Para, local} for several temperatures. 
	We observe no significant response as a function of the applied field even close to $T_M$ or above the spin-flop field. 
	%
	%
	\begin{figure}[ht!]
		\includegraphics[width = 8cm]{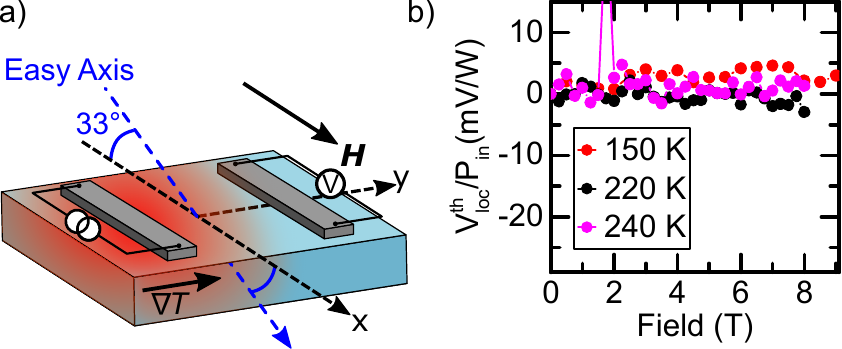}
		\caption{\label{fig:Wires Para, local} a) Schematic geometry of Pt wires relative to the easy-axis direction along $\vec{x}$. b) Local spin Seebeck signal for a voltage dropped along $\vec{x}$ as a function of the applied magnetic field, also along $\vec{x}$ for several temperatures.  }
	\end{figure}
	%
	%
		
	\begin{acknowledgments}
		A.R. and M.K. acknowledge support from the Graduate School of Excellence Materials Science in Mainz (DFG/GSC 266). A.R. and M.K. also acknowledge support from both MaHoJeRo (DAAD Spintronics network, project numbers 57334897 and 57524834) and SPIN+X (DFG SFB TRR 173, projects A01 and B02) and KAUST (OSR-2019-CRG8-4048.2). This work was supported by the Max Planck Graduate Center with the Johannes Gutenberg-Universität Mainz (MPGC). A. R., R. L., M.E., U. N. and M.K. acknowledge support from the DFG project number 423441604. R.L. acknowledges the European Union’s Horizon 2020 research and innovation programme under the Marie Skłodowska-Curie grant agreement FAST number 752195. This work has received funding from the European Union's Horizon 2020 research and innovation programme under Grant Agreement No. 863155 (s-Nebula). L.S. acknowledges support by a Mercator fellowship of the DFG. A.D. and L.S. acknowledge the support provided by the Ministry of Innovation and the National Research, Development, and Innovation (NRDI) Office under Projects No.\ PD124380, PD134579, K131938 and through the NRDI Fund (TKP2020 IES, Grant No.\ BME-IE-NAT). Computing resources for the \emph{ab initio} calculations were provided by Governmental Information Technology Development Agency's (KIFÜ) cluster in Debrecen, Hungary.
	\end{acknowledgments}

	%
	
\end{document}